\documentclass{IEEEtran}
\IEEEoverridecommandlockouts
\usepackage{cite}
\usepackage{amsmath,amssymb,amsfonts}
\usepackage{algorithmic}
\usepackage{graphicx}
\usepackage{textcomp}
\usepackage{xcolor}
\usepackage{setspace}
\def\BibTeX{{\rm B\kern-.05em{\sc i\kern-.025em b}\kern-.08emT\kern-.1667em\lower.7ex\hbox{E}\kern-.125emX}}
\usepackage{subfigure}

\usepackage{enumitem}

\title{Fixed-Gain AF Relaying for  RF-THz Wireless System  over $\alpha$-$\kappa$-$\mu$ Shadowed and $\alpha$-$\mu$ Channels}

\author{Pranay Bhardwaj,~\IEEEmembership{Graduate Student Member,~IEEE} and S.~M.~ Zafaruddin,~\IEEEmembership{Senior Member,~IEEE}
	\thanks{This work was supported in part by the Science and Engineering Research Board (SERB), Department of Science and Technology (DST), Government of India, under Start-up Research Grant SRG/2019/002345.}
	\thanks{Pranay Bhardwaj (p20200026@pilani.bits-pilani.ac.in) and S.~M.~Zafaruddin (syed.zafaruddin@pilani.bits-pilani.ac.in)  are  with  the Department of Electrical and Electronics Engineering, Birla Institute of Technology and Science, Pilani, Pilani-333031, Rajasthan, India.}
}

\begin{document}
	\maketitle 
\begin{abstract}
	Recent research investigates the decode-and-forward (DF) relaying for mixed radio frequency (RF) and terahertz (THz) wireless links  with zero-boresight pointing errors. In this letter, we analyze the performance of a fixed-gain amplify-and-forward (AF) relaying for the RF-THz link to interface the access network on the RF technology with wireless  THz transmissions. We develop probability density function (PDF) and cumulative distribution function (CDF) of the end-to-end SNR for the relay-assisted system in terms of bivariate Fox's H function considering  $\alpha$-$\mu$ fading for  the THz system with non-zero boresight pointing errors and $\alpha$-$\kappa$-$\mu$ shadowed ($\alpha$-KMS) fading model for the RF link. Using the derived PDF and CDF, we present exact analytical expressions of the outage probability, average bit-error-rate (BER), and ergodic capacity of the considered system. We also analyze the outage probability and average BER asymptotically for a better insight into the system behavior at high SNR. We use  simulations to compare the performance of the AF relaying having a semi-blind gain factor with the recently proposed DF relaying for THz-RF transmissions.
\end{abstract}		

\section{Introduction}
Terahertz (THz) communication is an emerging  technology for backhaul/fronthaul applications for next-generation wireless networks \cite{Elayan_2019, Koenig_2013_nature}. The THz link is less susceptible to the signal interference and can provide enormous unlicensed bandwidth to support high capacity links for broadband access in small cells and cell-free networks. However, the THz signal transmissions behave differently from the conventional radio frequency (RF) since the THz link is subjected to  random pointing errors caused by the misalignment between transmitter and receiver antenna beams and incurs hardware impairments at higher frequencies in addition to the signal fading and path loss \cite{KOKKONIEMI2020}. At the physical layer, an integration of line-of-sight THz transmissions for fronthauling and  radio frequency (RF) connectivity for the end-users can be a viable system configuration for 6G wireless networks, especially  in difficult terrains.

Cooperative relaying is an efficient technique to increase the data rate and extend the coverage range of wireless transmissions.  The use of relaying at THz frequencies has recently been studied   \cite{ Xia_2017,Giorgos2020, Boulogeorgos_2020_THz_THz,huang_2021_multihop_RIS_THz, Boulogeorgos_Error,Pranay_2021_TVT, Rong_2017, Abbasi_2017,Mir2020}.   The authors in \cite{Boulogeorgos_2020_THz_THz}  presented  the outage probability of a dual-hop  THz-THz link using the decode-and-forward (DF) relaying protocol. Recently, the authors in \cite{Boulogeorgos_Error,Pranay_2021_TVT} considered the DF relaying protocol to facilitate  data transmissions between the THz-RF mixed link. They have developed outage probability, average bit-error-rate (BER), and ergodic capacity  performance for  THz-RF transmissions by deriving  probability density function (PDF) and cumulative distribution function (CDF) of the THz link in terms of incomplete Gamma function over  $\alpha$-$\mu$ fading with zero-boresight pointing errors.   It is a well-known fact that the fixed-gain amplify-and-forward (AF) relaying possesses desirable characteristics of lower computational complexity and does not require continuous monitoring of the  channel state information (CSI) for decoding at the relay \cite{Hasna_2004_AF}.  There is limited research on  AF relaying for THz systems. The authors in \cite{Rong_2017} considered an AF relay  for nano-scale THz transmissions without considering the effect of short-term fading. 
Considering  Rayleigh fading, \cite{Abbasi_2017} studied an AF-assisted cooperative In-Vivo nano communication   at THz frequencies. The authors in  \cite{Liang_2021_arxiv}  analyzed the THz-THz dual-hop system with fixed-gain relaying considering zero-boresight pointing errors. It should be mentioned that the AF relaying has been extensively studied for various wireless technologies  such as RF-RF \cite{Hasna_2004_AF,ALVI_2019}, RF-free space optics (FSO) \cite{Lee_2011_FSO_RF,Ashrafzadeh_2019_FSO_RF}, RF-power line communications (PLC) \cite{Yang2021_plc},  mmWave-FSO\cite{Trigui_2019_mmWave_FSO,Zhang_2020_mmWave_FSO}, PLC-visible light communications (VLC) \cite{Gheth_2018_PLC_VLC}, and RF-underwater wireless communications (UWOC) \cite{Li2020}.

In this letter, we analyze the performance of a mixed RF-THz wireless system assisted by a fixed-gain AF relaying by considering  non-zero boresight pointing errors   with $\alpha$-$\mu$ fading for the THz and  generalized $\alpha$-$\kappa$-$\mu$ shadowed ($\alpha$-KMS) fading \cite{Espinosa_2019_alpha_kms} for the RF link.  To the best of the authors' knowledge, 		generalized pointing errors has not yet been considered for the THz link and  the use of $\alpha$-KMS has not been studied for the dual-hop relaying for mixed systems.  It should be mentioned that a recent measurement campaign validates $\alpha$-$\mu$ distribution for the short-term fading at $152$ \mbox{GHz} for a link length within $50$ \mbox{m} \cite{Papasotiriou2021}. We list the major contributions of the paper as follows:
\begin{itemize}[leftmargin=*]
	\item We provide statistical results on the signal-to-noise ratio (SNR)  by  deriving PDF and CDF of the THz link under the  joint effect of deterministic 
	path-loss,  $\alpha$-$\mu$ short-term fading,   and generalized pointing errors model. The derived PDF and CDF are also valid for real-valued $\alpha$ and $\mu$ parameters  and are represented in terms of Meijer's G function for an  elegant performance analysis  with the zero-boresight model as a special case.
	\item Using the derived PDF and CDF of the THz link, we derive novel density and distribution functions of the end-to-end  SNR for the RF-THz mixed link integrated with a fixed-gain AF relay. 
	\item We develop  exact analytical expressions of outage probability, average BER, and ergodic capacity of  the relay-assisted system in terms of bivariate Fox's function. By computing residues of Fox's H function at each pole, we also provide  asymptotic expressions  for the outage probability and the average BER at high SNR in terms of simpler Gamma functions, and derive the diversity order of the system.
\end{itemize}

\section{System Model}\label{sec:system_model}
We consider a mixed fronthaul/radio access system for uplink data transmissions where a fixed-gain AF relay facilitates communication between the source and destination. We establish broadband radio access from the source to the relay over RF and fronthaul link from the relay to the destination  over THz. The relay includes a frequency up-converter to generate  THz signals from a low-frequency RF. There is no direct link between the source and destination since both THz and RF operate on different carrier frequencies. The THz link is subjected to pointing errors in addition to the path gain, short-term fading, and hardware impairments of the transmitter and receiver. 

In the first hop, the received signal $y_R$\footnote{\emph{Notations}:  Subscripts $(\cdot)_{R}$,  $(\cdot)_D$, $(\cdot)_{r}$,  and $(\cdot)_t$ denote the  relay,  destination, the first link RF,  and second link THz, respectively. $G_{p,q}^{m,n}(.|.)$ and $H_{p,q}^{m,n}(.|.)$ denotes Meijer's G and Fox's H-functions, respectively.} 
 at the relay is expressed as $y_R = H_{r}  h_{fr}S + w_r$, where $S$ is the transmitted signal from the source, $w_r$ is the additive white Gaussian noise (AWGN) signal with variance $\sigma_{w_1}^2$, $H_{r}$ is the RF channel path-gain, and $h_{fr}$ is  the fading coefficient.  We  use the  generalized $\alpha$-$\kappa$-$\mu$ shadowed (a.k.a $\alpha$-KMS) distribution to model the short term fading  $|h_{fr}|$  for  the RF link with PDF \cite{Espinosa_2019_alpha_kms}:
\begin{eqnarray} \label{eq:pdf_alpha_kms}
	&f_{|{h_{f_r}}|}(x) = \frac{m_r^{m_r} \alpha_r}{2c^\mu_r \Gamma(\mu_r)(\mu_r \kappa_r+m_r)^m_r \bar{\gamma}_{r}} \big(\frac{x}{\bar{\gamma}_{r}}\big)^{\frac{\alpha_r\mu_r} {2}-1}  \nonumber \\ & {\rm exp}\big(-\frac{1}{c} \big(\frac{x}{\bar{\gamma}_{r}}\big)^{\frac{\alpha_r}{2}} \big) {}_1F_1 \big(m_r, \mu_r; \frac{\mu_r \kappa_r}{c(\mu_r \kappa_r+m_r)} \big(\frac{x}{\bar{\gamma}_{r}}\big)^{\frac{\alpha_r}{2}}  \big)
\end{eqnarray}
where $\{\alpha_r, \kappa_r, \mu_r, m_r\}$ are the fading parameters,  $ {}_1F_1 $ is the confluent Hypergeometric function and the parameter $c$ is defined in \cite{Espinosa_2019_alpha_kms}. In the second hop, the relay amplifies the incoming signal  $y_R$ with  a gain ${\mathcal{G}}$ to get the received signal at the destination (assuming negligible hardware distortions \cite{Boulogeorgos_Error,Pranay_2021_TVT}) $ y_D = {H_{t}}h_p h_{ft}  {\mathcal{G}} y_R + w_t$, where $ H_{t} $ is the path gain of THz link,   $w_t$ is the AWGN with variance $\sigma_{w_t}^2$, $h_p$ models pointing errors, and  $|h_{ft}|$ is the  short-term  fading of the THz link with PDF:
\begin{eqnarray} \label{eqn:pdf_hf_thz}
&	f_{|h_{ft}|}(x) = \frac{\alpha \mu_t^{\mu_t}}{ \Omega^{\alpha_t\mu_t}\Gamma (\mu_t)} x^{\alpha_t\mu_t-1} \exp \big(- \frac {\mu_t}  {\Omega^{\alpha_t\mu_t}}  {x^{\alpha_t}}\big)
\end{eqnarray}
where $\{\alpha_t, \mu_t, \Omega\}$ are the fading parameters for the THz link, and $\Gamma (\cdot)$ denotes the Gamma function.  We use the generalized non-zero boresight statistical model for $h_{p}$ \cite{Yang2014}:
\begin{eqnarray}\label{eq:pdf_hp}
	&f_{h_p}(h_p) = \frac{\phi\exp\big(\frac{-s^2}{2\sigma^2}\big)}{A_{0}^{\phi}}h_{p}^{\phi-1} I_0\bigg(\frac{s}{\sigma^2}\sqrt{\frac{w^2_{z_{\rm{eq}}}\ln \frac{S_0}{h_p}}{2}}\bigg)
\end{eqnarray}
where  $s=\sqrt{\mu_x^2+\mu_y^2}$ is the boresight displacement with $\mu_x$ and $\mu_y$ representing  horizontal and vertical boresight values, respectively,  $S_0$ is the fraction of collected power without pointing errors, $\phi$ is the ratio of normalized beam-width to the jitter,   and $I_0(\cdot)$ denotes  the modified Bessel function of the first kind with order zero.

We denote by $A_t = \frac{\alpha_t\mu_t^{\mu_t}}{\Omega^{\alpha_t\mu_t} \Gamma(\mu_t)}$, and $B_t = \frac{\mu_t}{\Omega^{\alpha_t\mu_t}}$.  Denoting $\bar{\gamma}_r= \frac{P_r |H_{r}|^2}{\sigma_{w_r}^2}$ with $P_r$ as the transmit power at the source for the RF transmission and  $\bar{\gamma}_t= \frac{P_t |H_{t}|^2}{\sigma_{w_t}^2}$  with $P_t$ as the transmit power at the relay for the THz transmission, we express the  SNR   of  the RF link as 	$ \gamma_r=\bar{\gamma}_r|h_f|^2$ and  the  SNR of the THz as $\gamma_t=\bar{\gamma}_t|h_fh_p|^2$. The end-to-end SNR AF relaying system is given by  $\gamma = \frac{\gamma_{r}\gamma_{t}}{\gamma_{t}+C}$ \cite{Hasna_2004_AF}
where $C={P_t}/{\mathcal{G}}^2\sigma_{w_t}^2$. For the blind AF relaying, an arbitrary value of ${\mathcal{G}}$
can be selected. However, in a semi-blind approach the fixed gain relaying factor ${\mathcal{G}}$ can be obtained using statistics of received signal of the first hop  	$C =(\mathbb{E}_{\gamma_r}(1+\gamma)^{-1})^{-1}$\cite{Hasna_2004_AF},  where $\mathbb{E}_{\gamma_r}$ denotes the expectation operator over the random variable $\gamma_r$.
Hence,  the well-known PDF of end-to-end SNR $\gamma$ of the fixed-gain  AF relayed system is given by
\begin{eqnarray} \label{eq:pdf_eqn_af}
	&f_\gamma(z) = \int_{0}^{\infty}  {f_{\gamma_r}\Big(\frac{z(x + C)}{x}\Big)} {f_{\gamma_t}(x)} \frac{x + C}{{x}} {dx}
\end{eqnarray} 
\section{Performance Analysis}\label{sec:perf_analysis}	
In this section, we provide statistical results for the  AF relaying by  deriving   analytical expressions of the PDF and CDF of the THz link.  Using the derived statistical results, we analyze the outage probability, average BER, and ergodic capacity performance of mixed RF-THz system.
\subsection{Density and Distribution Functions}
In the following, we present the PDF and CDF of  SNR for the  THz link subjected to short-term fading and pointing errors. 
Using the limits of   $|h_{ft}|$ and $h_p$ in \eqref{eqn:pdf_hf_thz} and \eqref{eq:pdf_hp}, respectively, the PDF of $|h_{fp}|=h_p |h_{ft}|$ can represented as
\begin{eqnarray} \label{eq:combined_pdf_eqn}
	&f_{|h_{fp}|}(h) = \int _{0}^{S_0} \frac {1}{h_p} f_{h_{f}}\left ({\frac {h}{h_p}}\right) f_{|h_{p}|}(h_p) {\rm d}h_p 
\end{eqnarray}	
Using \eqref{eq:pdf_hp} with the series expansion  $I_0(x)=\sum_{k=0}^{\infty}\frac{\left(\frac{x}{2}\right)^{2k}}{(k!)^2}$   in \eqref{eq:combined_pdf_eqn} and utilizing the integral form of Meijer's G-function:
\begin{eqnarray} \label{eq:pdf_intermediate_eqn}
	&f_{h_{fp}}(h)= \frac{ A_t \phi\exp\left(\frac{-s^2}{2\sigma^2}\right) h^{{(\alpha_t\mu_t-1)}}}{A_{0}^{\phi}} \sum_{j=0}^{\infty}\frac{1}{(j!)^2} \left(\frac{s^2w^2_{z_{\rm{eq}}}}{8\sigma^4}\right)^{j}  \nonumber \\ & \frac{1}{2\pi i} \int_{\mathcal{L}}^{} \Gamma(0-u_1) ( B_t h^{\alpha_t })^{u} du  I_{1}
\end{eqnarray}
where  $ I_{1} = \int_{0}^{S_0} h_p^{(\phi-\alpha\mu-1)} h_p^{-\alpha_t u} \big(\ln \frac{S_0}{h_p}\big)^{j}{\rm d}h_p$. Substituting $\ln \frac{S_0}{h_p}=t $, we obtain $I_1=S_0^{(\phi-\alpha_t\mu_t+1-\alpha_t u)} \Gamma(j+1) \Big(\frac{\Gamma(1+\phi-\alpha\mu-\alpha u)} {\Gamma(\phi-\alpha\mu-\alpha u)}\Big)^{-(j+1)}$. Further, using $I_{1}$ in \eqref{eq:pdf_intermediate_eqn} and applying the definition of Fox's H-function \cite{Kilbas_2004} with a transformation  of the random variable  $f_{\gamma_t}(\gamma) = \frac{1}{2\sqrt{\gamma\bar{\gamma}_t}} f_{h_{pf}}\Big(\sqrt{\frac{\gamma}{\bar{\gamma}_t}}\Big) $ we get the PDF
\begin{eqnarray}	\label{eq:pdf_combined_hfp}
	&f_{\gamma_t}(\gamma)=  \frac{A_t \phi\exp\left(\frac{-s^2}{2\sigma^2}\right) S_0^{(-\alpha_t\mu_t+1)}  {\gamma}^{\frac{\alpha_t\mu_t}{2}-1}}{2 {\bar{\gamma}_{t}}^{\frac{\alpha_t\mu_t}{2}}}  \sum_{j=0}^{\infty}\frac{1}{j!}  \nonumber \\ & \Big(\frac{s^2w^2_{z_{\rm{eq}}}}{8\sigma^4}\Big)^{j}       H_{j+1,k+2}^{k+2,0}\Bigg[\frac{B_t \gamma^{\frac{\alpha_t}{2}}}{S_0^{\alpha_t} {\bar{\gamma}_t}^{\frac{\alpha_t}{2}}} \Bigg| \begin{matrix} (1+\phi-\alpha\mu,1)^{j+1} \\   (0,1), (\phi-\alpha\mu,1)^{j+1}  \end{matrix} \Bigg]
\end{eqnarray}
We use the PDF \eqref{eq:pdf_combined_hfp} in $\int_{0}^{\gamma} f_{\gamma}(z) dz$ and simplify the integral using the Mellin Barnes integral representation of the Fox's H-function to get the CDF:
\begin{eqnarray}	\label{eq:cdf_hfp}
	&F_{\gamma_{t}}(x) =  \frac{A_t \phi\exp\left(\frac{-s^2}{2\sigma^2}\right) S_0^{(-\alpha_t\mu_t+1)}  {\gamma}^{\frac{\alpha_t\mu_t}{2}}}{2 {\bar{\gamma}_{t}}^{\frac{\alpha_t\mu_t}{2}}}  \sum_{j=0}^{\infty}\frac{1}{j!}   \Big(\frac{s^2w^2_{z_{\rm{eq}}}}{8\sigma^4}\Big)^{j}  \nonumber \\ &     H_{j+1,j+3}^{k+2,1}\Bigg[\frac{B_t \gamma^{\frac{\alpha_t}{2}}}{S_0^{\alpha_t} {\bar{\gamma}_t}^{\frac{\alpha_t}{2}}} \Bigg| \begin{matrix} \big(1-\frac{\alpha_t\mu_t}{2}, \frac{\alpha_t}{2}\big), (1+\phi-\alpha\mu,1)^{j+1} \\   (0,1), (\phi-\alpha\mu,1)^{j+1}, \big(-\frac{\alpha_t\mu_t}{2}, \frac{\alpha_t}{2}\big)  \end{matrix} \Bigg]                       
\end{eqnarray}

Note that the  use of Meijer's G and  Fox's H functions is common in the research fraternity and can be efficiently evaluated using  built-in functions available in computational software such as MATLAB and MATHEMATICA. We capitalize  results of \eqref{eq:pdf_combined_hfp} and \eqref{eq:cdf_hfp} to present the PDF of SNR for the AF relaying in terms of bivariate Fox's H function.

Representing exponential and hypergeometric functions of  \eqref{eq:pdf_alpha_kms} into Meijer's G-function with à transformation of random variable $\gamma_r=\bar{\gamma}_r|h_f|^2$, we get
\begin{eqnarray} \label{eq:pdf_alpha_kms2}
	&f_{\gamma_r}(x) = \frac{m_r^{m_r} \alpha_r}{2c^\mu_r \Gamma(\mu_r)(\mu\kappa_r+m_r)^{m_r} \bar{x}} \big(\frac{x}{\bar{\gamma}_{r}}\big)^{\frac{\alpha_r\mu_r} {2}-1}  \nonumber \\ & G_{0, 1}^{1,0}\Big( \frac{x^{\frac{\alpha_r}{2}} }{c \bar{\gamma}_{r}^{\frac{\alpha_r}{2}} }  \Big| \begin{matrix} - \\ 0 \end{matrix} \Big)  \frac{\Gamma(\mu_r)}{\Gamma(m_r)}   G_{1, 2}^{1,1}\bigg( \frac{-\mu_r \kappa_r x^{\frac{\alpha_r}{2}}}{c(\mu_r \kappa_r+m_r) {\bar{\gamma}_{r}}^{\frac{\alpha}{2}}}  \Bigg| \begin{matrix} 1-m_r \\ 0, 1-\mu_r \end{matrix} \bigg)
\end{eqnarray}
Substituting \eqref{eq:pdf_combined_hfp} and \eqref{eq:pdf_alpha_kms2} in  \eqref{eq:pdf_eqn_af} and utilizing the integral representation of Fox's H-function with a change in the order of integration:  
\begin{eqnarray} \label{eq:pdf_proof_int}
	&f_\gamma (\gamma) =  \frac{m_r^{m_r} \alpha_r {\gamma}^{\frac{\alpha_r\mu_r}{2}-1}}{2c^{\mu_r}(\mu_r \kappa_r+m_r)^{m_r} \bar{\gamma}_{r} \Gamma(m_r) {\bar{\gamma}_{r}}^{\frac{\alpha_r\mu_r} {2}-1}}    \nonumber \\ & \frac{A_t \phi\exp\left(\frac{-s^2}{2\sigma^2}\right) S_0^{(-\alpha_t\mu_t+1)} }{2 {\bar{\gamma}_{t}}^{\frac{\alpha_t\mu_t}{2}}}  \sum_{j=0}^{\infty}\frac{\Gamma(j+1)}{(j!)^2}   \left(\frac{s^2w^2_{z_{\rm{eq}}}}{8\sigma^4}\right)^{j}   \nonumber \\ & \frac{1}{2\pi i} \int_{{\mathcal{L}_1}} \frac{\Gamma(0-u_1) \Gamma(0-u_1) \Gamma(m_r+u_1)}{\Gamma(\mu_r+u_1)} \bigg( \frac{-\mu_r \kappa_r {\gamma}^{\alpha_r}}{c^2(\mu_r \kappa_r+m_r) {\bar{\gamma}_{r}}^{\alpha_r}} \bigg)^{u_1} du_1 \nonumber \\ &  \frac{1}{2\pi i} \int_{{\mathcal{L}_2}}^{}   \frac{ \Gamma(0-u_2) \Gamma(\phi-\alpha_t\mu_t-\alpha u_2)^{(j+1)}} {\Gamma(1+\phi-\alpha_t\mu_t-\alpha_t u_2)^{(j+1)}} \Big( \frac{B_t}{S_0^{\alpha_t} {\bar{\gamma}_t}^{\frac{\alpha_t}{2}}}  \Big)^{u_2} du_2 I_2
\end{eqnarray}
where ${{\cal{L}}_1}$ and ${{\cal{L}}_2}$ denote the contour integrals. We use  \cite[(3.194/3)]{Gradshteyn} and \cite[(8.384/1)]{Gradshteyn} to solve the inner integral $I_2$ in terms of the compatible Gamma function:
\begin{eqnarray} \label{inner_int_pdf}
	&\int_{0}^{\infty} \Big(\frac{\gamma+C}{x}\Big)^{\big(\frac{\alpha_r\mu_r}{2}+\alpha_r u_1\big)} {\gamma}^{\frac{\alpha_t\mu_t+\alpha_t u_2}{2}-1}  d\gamma = \nonumber \\ & \frac{C^{\frac{\alpha_t\mu_t+\alpha_tu_2}{2}} \Gamma(\frac{-\alpha_t\mu_t-\alpha_t u_2}{2}) \Gamma(\frac{-\alpha_r\mu_r-2\alpha_r u_1 + \alpha_t\mu_t + \alpha_t u_2}{2})  }{\Gamma(\frac{-\alpha_r\mu_r -2\alpha_r u_1}{2})}
\end{eqnarray}
Finally, we substitute \eqref{inner_int_pdf} in \eqref{eq:pdf_proof_int}, and  to apply the definition of Fox's H function \cite[(1.1)]{Mittal_1972}, we use  $u_1 \to -u_1 $ in \eqref{eq:pdf_proof_int} to get the PDF for fixed-gain relaying:
\begin{eqnarray} \label{eq:pdf_fg}
	&f_\gamma (\gamma)\hspace{-0.5mm} =  \hspace{-0.5mm}  \frac{m_r^{m_r} \alpha_r  C^{\frac{\alpha_t\mu_t}{2}}  A_t \phi\exp\left(\frac{-s^2}{2\sigma^2}\right) S_0^{(-{\alpha_t}{\mu_t}+1)} {\gamma}^{\frac{\alpha_r\mu_r}{2}-1} }{4c^{\mu_r}(\mu_r \kappa_r+m_r)^{m_r}  \Gamma(m_r) {\bar{\gamma}_{r}}^{\frac{\alpha_r\mu_r} {2}} {\bar{\gamma}_{t}}^{\frac{\alpha_t\mu_t} {2}}} \hspace{-1mm}\sum_{j=0}^{\infty}  \nonumber \\ &   \frac{1}{j!} \hspace{-0.5mm} \bigg(\hspace{-1mm}\frac{s^2w^2_{z_{\rm{eq}}}}{8\sigma^4}\hspace{-1.5mm}\bigg)^{\hspace{-1mm}j}  \hspace{-1mm}  H_{1,0:3,2:j+1,j+3}^{0,1:1,2:j+3,0} \Bigg[\hspace{-1mm} \frac{ c^2(\mu_r \kappa_r+m_r) {\bar{\gamma}_{r}}^{\alpha_r}}{-\mu_r \kappa_r {\gamma}^{\alpha_r}} ; \frac{B_t C^{\frac{\alpha_t}{2}}}{S_0^{\alpha_t} {\bar{\gamma}_t}^{\frac{\alpha_t}{2}}} \Bigg| \begin{matrix}  V_1     \\   V_2  \end{matrix}  \Bigg]     
\end{eqnarray}
where $V_1\hspace{-0.5mm} = \hspace{-0.5mm}\bigl\{\big(1-\frac{\alpha_t\mu_t-\alpha_r\mu_r}{2}; \alpha_r, \frac{\alpha_t}{2}\big)\bigr\}; \bigl\{\big( 1,1 \big), \big( 1,1 \big),\big( \mu_r,1 \big)  \bigr\} \\ ; \bigl\{ \big(1+\phi-\alpha_t\mu_t, \alpha_t\big)^{j+1}  \bigr\}$, and $ V_2 = \bigl\{-\bigr\}; \bigl\{\big(m_r,1\big), \big( 1 +\frac{\alpha_r\mu_r}{2},\alpha_r \big)  \bigr\} ; \bigl\{ \big(0,1\big), \big(\phi-\alpha_t\mu_t, \alpha_t\big)^{j+1}, \big(-\frac{\alpha_t\mu_t}{2},\frac{\alpha_t}{2}\big)$.

\subsection{Outage Probability}
Outage probability is defined as the probability of SNR being less than a threshold value $\gamma_{th}$ i.e., $ P_{\rm out} = P(\gamma <\gamma_{th})=F_{\gamma}(\gamma_{\rm th})$. Thus, using \eqref{eq:pdf_fg} in $F_{\gamma}(\gamma_{}) = \int_{0}^{\gamma_{}} f_{\gamma}(z) {dz}$, and applying the definition of Fox's H function with the  following inner integral 
\begin{eqnarray}
	\int_{0}^{\gamma} {z}^{\frac{\alpha_r\mu_r}{2}-1} z^{-\alpha_r u_1} dz=\frac{z^{\frac{\alpha_r\mu_r}{2}} z^{{- \alpha_r u_1}} \Gamma\big(\frac{\alpha_r\mu_r-2 \alpha_r u_1}{2}\big)} {\Gamma\big(1+\frac{\alpha_r\mu_r-2 \alpha_r u_1}{2}\big)}
\end{eqnarray}
we get the CDF as 
\begin{eqnarray} \label{eq:cdf_fg}
&F_\gamma (\gamma)\hspace{-0.5mm} = \hspace{-0.5mm} \frac{m_r^{m_r} \alpha_r  C^{\frac{\alpha_t\mu_t}{2}}  A_t \phi\exp\left(\frac{-s^2}{2\sigma^2}\right) S_0^{(-{\alpha_t}{\mu_t}+1)} {\gamma}^{\frac{\alpha_r\mu_r}{2}} }{4c^{\mu_r}(\mu_r \kappa_r+m_r)^{m_r}  \Gamma(m_r) {\bar{\gamma}_{r}}^{\frac{\alpha_r\mu_r} {2}} {\bar{\gamma}_{t}}^{\frac{\alpha_t\mu_t} {2}}} \hspace{-1mm} \sum_{j=0}^{\infty} \frac{1}{j!}    \nonumber \\ & \bigg(\hspace{-1mm}\frac{s^2w^2_{z_{\rm{eq}}}}{8\sigma^4}\hspace{-1mm}\bigg)^{\hspace{-1mm}j} \hspace{-1mm} H_{1,0:4,3:j+1,j+3}^{0,1:2,2:j+3,0} \Bigg[ \frac{ c^2(\mu_r \kappa_r+m_r) {\bar{\gamma}_{r}}^{\alpha_r}}{-\mu_r \kappa_r {\gamma}^{\alpha_r}} ; \frac{B_t C^{\frac{\alpha_t}{2}}}{S_0^{\alpha_t} {\bar{\gamma}_t}^{\frac{\alpha_t}{2}}} \Bigg| \begin{matrix}   Q_1    \\    Q_2 	\end{matrix}  \Bigg]     
\end{eqnarray}
where $Q_1 = \bigl\{\big(1-\frac{\alpha_t\mu_t-\alpha_r\mu_r}{2}; \alpha_r, \frac{\alpha_t}{2}\big)\bigr\}; \bigl\{\big( 1,1 \big), \big( 1,1 \big),\\ \big( \mu_r,1 \big),\big(1+\frac{\alpha_r\mu_r}{2}, \alpha_r\big)  \bigr\} ; \bigl\{ \big(1+\phi-\alpha_t\mu_t, \alpha_t\big)^{j+1}  \bigr\} $, and $ Q_2 = \bigl\{-\bigr\}; \bigl\{\big(m_r,1\big), \big(\frac{\alpha_r\mu_r}{2}, \alpha_r\big), \big( 1 +\frac{\alpha_r\mu_r}{2},\alpha_r \big)  \bigr\} ; \bigl\{ \big(0,1\big), \big(\phi-\alpha_t\mu_t, \alpha_t\big)^{j+1}, \big(-\frac{\alpha_t\mu_t}{2},\frac{\alpha_t}{2}\big)   \bigr\}$.

To derive the asymptotic  outage probability in the high SNR regime, we use  \cite[Theorems 1.7, 1.11]{Kilbas_2004} and compute  residues of \eqref{eq:cdf_fg} for both contours ${\cal{L}}_1$ and ${\cal{L}}_2$ at poles $u_1=0,0$, $\frac{-\alpha_r\mu_r  + \alpha_t\mu_t+ \alpha_tu_2}{2\alpha_r}$ and $s_2=0$, $-\mu_t$, and $\frac{\phi-\alpha_t\mu_t}{\alpha_t}$. Simplifying the derived residues, we present the asymptotic expression in \eqref{eq:outage_ber_asymptotic}. It should be emphasized that the consideration of all the poles may result into our asymptotic analysis close to the exact  for a wide range of SNR.  Further,  it can be easily seen from \eqref{eq:outage_ber_asymptotic} that the outage  diversity order of the system is $\min \big\{ \frac{\alpha_r \mu_r}{2},\frac{\alpha_t \mu_t}{2}, \frac{\phi}{2}\big\}$. Note that  the derived  diversity order for the THz-RF can be confirmed individually with previous results on $\alpha$-$\mu$ fading \cite{Pranay_2021_TVT}  and $\alpha$-KMS \cite{Espinosa_2019_alpha_kms}.

\begin{figure*}
	\small
\begin{eqnarray}\label{eq:outage_ber_asymptotic}
	\Psi^\infty = G_0 \Bigg[G_1  \bigg( \frac{-\mu_r \kappa_r \gamma^{\alpha_r} }{ c^2(\mu_r \kappa_r+m_r) } \bigg)^{\frac{ \alpha_t\mu_t}{2\alpha_r}-\frac{\mu_r}{2}}    {\bar{\gamma}_{r}}^{-\frac{ \alpha_t\mu_t}{2}}  +   G_2  \Big( \frac{B_t C^{\frac{\alpha_t}{2}}}{S_0^{\alpha_t} {\bar{\gamma}_t}^{\frac{\alpha_t}{2}}} \Big)^{\hspace{-1mm}\frac{\phi-\alpha_t\mu_t}{\alpha_t}}        \bigg(\hspace{-1mm} \frac{-\mu_r \kappa_r \gamma^{\alpha_r} }{ c^2(\mu_r \kappa_r+m_r) } \hspace{-1mm}\bigg)^{\hspace{-1mm}\frac{\phi}{2\alpha_r}-\frac{\mu_r}{2}}   {\bar{\gamma}_{r}}^{-\frac{\phi}{2}}   +  G_3 {\bar{\gamma}_{r}}^{-\frac{\alpha_r\mu_r} {2}}   \Bigg]
\end{eqnarray}
where $	G_0=\frac{m_r^{m_r} \alpha_r   A_t \phi\exp\left(\frac{-s^2}{2\sigma^2}\right) S_0^{(-{\alpha_t}{\mu_t}+1)}  \varphi }{4c^{\mu_r}(\mu_r \kappa_r+m_r)^{m_r}  \Gamma(m_r)}  \sum_{j=0}^{\infty}\frac{1}{j!}   \left(\frac{s^2w^2_{z_{\rm{eq}}}}{8\sigma^4}\right)^{j}$, $G_1=\frac{\Gamma(\frac{\alpha_r\mu_r-\alpha_t\mu_t }{2\alpha_r}) \Gamma(\frac{\alpha_r\mu_r - \alpha_t\mu_t }{2\alpha_r}) \Gamma(m_r+\frac{-\alpha_r\mu_r  + \alpha_t\mu_t}{2\alpha_r}) \Gamma(\frac{-\alpha_t\mu_t}{2})C^{\frac{\alpha_t\mu_t}{2}}\zeta_1}{\Gamma(\mu_r+\frac{-\alpha_r\mu_r  + \alpha_t\mu_t}{2\alpha_r}) \Gamma(-\alpha_t\mu_t) \alpha_t\mu_t (\phi-\alpha_t\mu_t)^{(j+1)}  {\bar{\gamma}_{t}}^{\frac{\alpha_t\mu_t} {2}}} $

$ G_2 = \frac{\Gamma(\frac{\alpha_r\mu_r - \phi}{2\alpha_r}) \Gamma(\frac{\alpha_r\mu_r - \phi}{2\alpha_r}) \Gamma(m_r+\frac{-\alpha_r\mu_r  \phi}{2\alpha_r})  \Gamma(\frac{\alpha_t\mu_t-\phi}{\alpha_t}) \Gamma(\frac{-\phi}{2}) C^{\frac{\alpha_t\mu_t} {2}} \zeta_2}{\Gamma(\mu_r+\frac{-\alpha_r\mu_r  +\phi}{2\alpha_r}) \Gamma(-\phi) \phi  {\bar{\gamma}_{t}}^{\frac{\alpha_t\mu_t} {2}}} $ $ G_3 = \Big(\frac{2 C^{\frac{\alpha_t\mu_t} {2}}\zeta_3}{ {\bar{\gamma}_{t}}^{\frac{\alpha_t\mu_t} {2}}}\Big)    \Bigg(  \frac{\Gamma(\frac{-\alpha_r\mu_r  + \alpha_t\mu_t}{2}) \Gamma(m_r) \Gamma(\frac{-\alpha_t\mu_t}{2})}{\Gamma(\mu_r) \Gamma(\frac{-\alpha_r\mu_r}{2}) \frac{\alpha_r\mu_r}{2} (\phi-\alpha_t\mu_t)^{(j+1)}}      +     \frac{\Gamma(m_r) \Gamma(\mu_t)}{\Gamma(\mu_r)  \frac{\alpha_r\mu_r}{2} (\phi)^{(j+1)}}  \Big( \frac{B_t C^{\frac{\alpha_t}{2}}}{S_0^{\alpha_t} {\bar{\gamma}_t}^{\frac{\alpha_t}{2}}} \Big)^{-\mu_t}           +        \frac{\Gamma(\frac{-\alpha_r\mu_r  + \phi}{2}) \Gamma(m_r) \Gamma(\frac{\alpha_t\mu_t-\phi}{\alpha_t}) \Gamma(\frac{-\phi}{2})}{\Gamma(\mu_r) \Gamma(\frac{-\alpha_r\mu_r}{2}) \frac{\alpha_r\mu_r}{2}}    \Big( \frac{B_t C^{\frac{\alpha_t}{2}}}{S_0^{\alpha_t} {\bar{\gamma}_t}^{\frac{\alpha_t}{2}}} \Big)^{\frac{\phi-\alpha_t\mu_t}{\alpha_t}} + \frac{\Gamma(\frac{\mu_r }{2}) \Gamma(m_r+\frac{-\mu_r }{2})  \Gamma(\mu_t) C^{\frac{\alpha_t\mu_t} {2}} \zeta_4}{(\phi)^{(j+1)} {\bar{\gamma}_{t}}^{\frac{\alpha_t\mu_t} {2}}} \bigg( \frac{-\mu_r \kappa_r \gamma^{\alpha_r} }{ c^2(\mu_r \kappa_r+m_r) } \bigg)^{\frac{-\mu_r }{2}}   \\   \Big( \frac{B_t C^{\frac{\alpha_t}{2}}}{S_0^{\alpha_t} {\bar{\gamma}_t}^{\frac{\alpha_t}{2}}} \Big)^{-\mu_t}   \Bigg) $;  $\Psi^\infty = P_{\rm out}^\infty$ when $\varphi = {\rm \gamma_{\rm th}}^{\frac{\alpha_t\mu_t}{2}}$ and $\zeta_1=\zeta_2=\zeta_3=\zeta_4=1$;  $\Psi^\infty = \overline{BER}^\infty$ when  $\varphi = \frac{{q}^{-(\frac{\alpha_r\mu_r}{2} + p)}}{2\Gamma(p)}$,	$\zeta_1 = \Gamma\big(p+\frac{\alpha_r\mu_r}{2}\big)$, $\zeta_2=\Gamma(p)$, $\zeta_3 = \Gamma\big(p+\frac{\alpha_t\mu_t}{2}\big)$, and $\zeta_4 = \Gamma\big(p+\frac{\phi}{2}\big) $.
	\hrule
\end{figure*}
\normalsize	
\subsection{Average BER}\label{sec:av_ber}
The average BER of a communication system is given as:
\begin{eqnarray} \label{eq:ber}
&\overline{BER} = \frac{q^p}{2\Gamma(p)}\int_{0}^{\infty} \gamma^{p-1} {e^{{-q \gamma}}} F_{\gamma} (\gamma)   d\gamma
\end{eqnarray}
where $p$ and $q$ are modulation-dependent constants.
Using CDF of \eqref{eq:cdf_fg} in \eqref{eq:ber}, the average BER can be expressed as
\begin{eqnarray} \label{eq:ber_proof_int}
	&\overline{BER} =    \frac{m_r^{m_r} \alpha_r  C^{\frac{\alpha_t\mu_t}{2}}  A_t \phi\exp\left(\frac{-s^2}{2\sigma^2}\right) S_0^{(-{\alpha_t}{\mu_t}+1)} q^p }{8c^{\mu_r}(\mu_r \kappa_r+m_r)^{m_r}  \Gamma(m_r) \Gamma(p) {\bar{\gamma}_{r}}^{\frac{\alpha_r\mu_r} {2}} {\bar{\gamma}_{t}}^{\frac{\alpha_t\mu_t} {2}}}     \nonumber \\ &    \sum_{j=0}^{\infty}\frac{\Gamma(j+1)}{(j!)^2}   \left(\frac{s^2w^2_{z_{\rm{eq}}}}{8\sigma^4}\right)^{j}   \nonumber \\ & \frac{1}{2\pi i} \int_{\mathcal{L}_1} \frac{\Gamma(0+u_1) \Gamma(0+u_1) \Gamma(m_r-u_1)}{\Gamma(\mu_r-u_1)} \bigg( \frac{ c^2(\mu_r \kappa_r+m_r) {\bar{\gamma}_{r}}^{\alpha_r}}{-\mu_r \kappa_r } \bigg)^{u_1} du_1 \nonumber \\ &  \frac{1}{2\pi i} \int_{\mathcal{L}_2}^{}   \frac{ \Gamma(0-u_2) \Gamma(\phi-\alpha_t\mu_t-\alpha_t u_2)^{(j+1)}} {\Gamma(1+\phi-\alpha_t\mu_t-\alpha_t u_2)^{(j+1)}} \Big( \frac{B_t C^{\frac{\alpha_t}{2}}}{S_0^{\alpha_t} {\bar{\gamma}_t}^{\frac{\alpha_t}{2}}} \Big)^{u_2} du_2 \nonumber \\ & \frac{ \Gamma(\frac{-\alpha_t\mu_t-\alpha_tu_2}{2}) \Gamma(\frac{-\alpha_r\mu_r  + \alpha_t\mu_t  + 2\alpha_ru_1 + \alpha_tu_2}{2})  }{\Gamma(\frac{-\alpha_r\mu_r +2\alpha_ru_1}{2})}  \frac{ \Gamma\big(\frac{\alpha_r\mu_r-2 \alpha_ru_1}{2}\big)} {\Gamma\big(1+\frac{\alpha_r\mu_r-2 \alpha_ru_1}{2}\big)} \nonumber \\ & \int_{0}^{\infty} \gamma^{p-1} e^{-q \gamma} \gamma^{\frac{\alpha_r\mu_r-2 \alpha_ru_1}{2}}   d\gamma
\end{eqnarray}
Using the solution of inner integral $\int_{0}^{\infty} \gamma^{p-1} e^{-q \gamma} \gamma^{\frac{\alpha_r\mu_r-2 \alpha_ru_1}{2}}   d\gamma$ \cite[(3.381/4)]{Gradshteyn} in terms of Gamma function, and applying the definition of Fox's H function \cite[(1.1)]{Mittal_1972}, we get 
\begin{eqnarray} \label{eq:ber_fg}
	&\overline{BER} \hspace{-0.5mm}= \hspace{-0.5mm} \frac{m_r^{\hspace{-0.5mm}m_r} \alpha_r  C^{\frac{\alpha_t\mu_t}{2}}  A_t \phi\exp\big(\hspace{-0.5mm}\frac{-s^2}{2\sigma^2}\hspace{-0.5mm}\big) S_0^{(-{\alpha_t}{\mu_t}+1)} q^{\hspace{-0.5mm}\frac{-\alpha_r\mu_r}{2}}}{8c^{\mu_r}(\mu_r \kappa_r+m_r)^{m_r}  \Gamma(m_r) \Gamma(p) {\bar{\gamma}_{r}}^{\frac{\alpha_r\mu_r} {2}} {\bar{\gamma}_{t}}^{\frac{\alpha_t\mu_t} {2}} }\hspace{-1mm} \sum_{j=0}^{\infty}\hspace{-1mm} \frac{1}{j!}    \nonumber \\ & \bigg(\hspace{-1.5mm}\frac{s^2w^2_{z_{\rm{eq}}}}{8\sigma^4}\hspace{-1mm}\bigg)^{\hspace{-1mm}j} \hspace{-1mm} H_{1,0:4,4:j+1,j+3}^{0,1:3,2:j+3,0} \hspace{-1mm}\Bigg[\hspace{-1mm} \frac{ c^2(\mu_r \kappa_r+m_r) {\bar{\gamma}_{r}}^{\alpha_r} q^{\alpha_r}}{-\mu_r \kappa_r} ;\hspace{-1mm} \frac{B_t C^{\frac{\alpha_t}{2}}}{S_0^{\alpha_t} {\bar{\gamma}_t}^{\frac{\alpha_t}{2}}} \Bigg| \begin{matrix}  U_1   \\  U_2  	\end{matrix}  \Bigg]     
\end{eqnarray}
where $U_1 \hspace{-1mm}  = \bigl\{\big(1-\frac{\alpha_t\mu_t-\alpha_r\mu_r}{2}; \alpha_r, \frac{\alpha_t}{2}\big)\bigr\}; \bigl\{\big( 1,1 \big), \big( 1,1 \big),\big( \mu_r,1 \big), \\ \big(1+\frac{\alpha_r\mu_r}{2}, \alpha_r\big)  \bigr\} ; \bigl\{ \big(1+\phi-\alpha_t\mu_t, \alpha_t\big)^{j+1}  \bigr\}$, and $U_2  = \bigl\{-\bigr\}; \bigl\{\big(m_r,1\big), \big(\frac{\alpha_r\mu_r}{2}, \alpha_r\big),\big(p+\frac{\alpha_r\mu_r}{2}, \alpha_r\big), \big( 1 +\frac{\alpha_r\mu_r}{2},\alpha_r \big)  \bigr\} ; \bigl\{ \big(0,1\big), \big(\phi-\alpha_t\mu_t, \alpha_t\big)^{j+1}, \big(-\frac{\alpha_t\mu_t}{2},\frac{\alpha_t}{2}\big) \bigr\}  $. 

Similar to the asymptotic expression of the outage probability, we derive the average BER at the high SNR  $\overline{BER}^{\infty}$ in terms of simpler Gamma function, as given in \eqref{eq:outage_ber_asymptotic}, and the corresponding diversity order as $\min \big\{ \frac{\alpha_r \mu_r}{2},\frac{\alpha_t \mu_t}{2}, \frac{\phi}{2}\big\}$.

\subsection{Ergodic Capacity}\label{sec:capacity}
The ergodic capacity is defined as
\begin{eqnarray} \label{eq:capacity_eqn}
&\overline{C}=  \int_{0}^{\infty} {\log_2}(1+\gamma) f_\gamma(\gamma) d\gamma 
\end{eqnarray}

Thus, substituting the PDF  \eqref{eq:pdf_fg} in \eqref{eq:capacity_eqn}, we apply the definition of Fox's H resulting into the inner integral as  \cite[(4.293.3)]{Gradshteyn}
\begin{eqnarray}\label{eq:zaf5}
	&\int_{0}^{\infty} \hspace{-0mm} {\rm log_2}(1+\gamma) {\gamma}^{\frac{\alpha_r\mu_r}{2}-1} \gamma^{-\alpha_ru_1} d\gamma = \hspace{-1mm} \frac{\pi Csc\big(\pi \frac{\alpha_r\mu_r-2\alpha_ru_1}{2}\big)}{ { \log}(2)\frac{\alpha_r\mu_r-2\alpha_ru_1}{2}}
\end{eqnarray}
Finally, we apply the identity $\pi Csc(\pi a) = \Gamma(a)\Gamma(1-a)$ in \eqref{eq:zaf5} along with the Fox's H function \cite [(1.1)]{Mittal_1972} to get the ergodic capacity:
\begin{eqnarray} \label{eq:capacity_fg}
	&\overline{C}\hspace{-0.5mm} = \hspace{-0.5mm} \frac{m_r^{m_r} \alpha_r  C^{\frac{\alpha_t\mu_t}{2}}  A_t \phi\exp\left(\frac{-s^2}{2\sigma^2}\right) S_0^{(-{\alpha_t}{\mu_t}+1)} }{{\rm log}(2)4c^{\mu_r}(\mu_r \kappa_r+m_r)^{m_r}  \Gamma(m_r) {\bar{\gamma}_{r}}^{\frac{\alpha_r\mu_r} {2}} {\bar{\gamma}_{t}}^{\frac{\alpha_t\mu_t} {2}}} \sum_{j=0}^{\infty} \hspace{-1mm} \frac{1}{j!}  \nonumber \\ & \bigg(\hspace{-1.5mm}\frac{s^2w^2_{z_{\rm{eq}}}}{8\sigma^4}\hspace{-1.5mm}\bigg)^{\hspace{-1mm}j}  \hspace{-1mm} H_{1,0:4,5:j+1,j+3}^{0,1:4,2:j+3,0} \Bigg[\hspace{-1mm} \frac{ c^2(\mu_r \kappa_r+m_r) {\bar{\gamma}_{r}}^{\alpha_r}}{-\mu_r \kappa_r} ;\hspace{-0.5mm} \frac{B_t C^{\frac{\alpha_t}{2}}}{S_0^{\alpha_t} {\bar{\gamma}_t}^{\frac{\alpha_t}{2}}} \Bigg| \begin{matrix}  D_1     \\  D_2   	\end{matrix}  \Bigg]     
\end{eqnarray}
where  $D_1\hspace{-1mm} = \bigl\{\big(1-\frac{\alpha_t\mu_t-\alpha_r\mu_r}{2}; \alpha_r, \frac{\alpha_t}{2}\big)\bigr\}; \bigl\{\big( 1,1 \big), \big( 1,1 \big),\big( \mu_r,1 \big), \\ \big(1+\frac{\alpha_r\mu_r}{2},\alpha_r\big)  \bigr\} ; \bigl\{ \big(1+\phi-\alpha_t\mu_t, \alpha_t\big)^{j+1}  \bigr\}$, and $D_2= \bigl\{-\bigr\}; \bigl\{\big(m_r,1\big), \big(\frac{\alpha_r\mu_r}{2},\alpha_r\big), \big(\frac{\alpha_r\mu_r}{2},\alpha_r\big), \big(1-\frac{\alpha_r\mu_r}{2},\alpha_r\big), \big( 1 +\frac{\alpha_r\mu_r}{2},\alpha_r \big)  \bigr\} ; \bigl\{ \big(0,1\big), \big(\phi-\alpha_t\mu_t, \alpha_t\big)^{j+1}, \big(-\frac{\alpha_t\mu_t}{2},\frac{\alpha_t}{2}\big)   \bigr\} $.

\begin{figure*}[tp]
		\subfigure[Outage probability at $\alpha_t=1.5$, $\gamma_{\rm th} = 4 \mbox{dB}$. ]{\includegraphics[width=6cm, height=4.1cm]{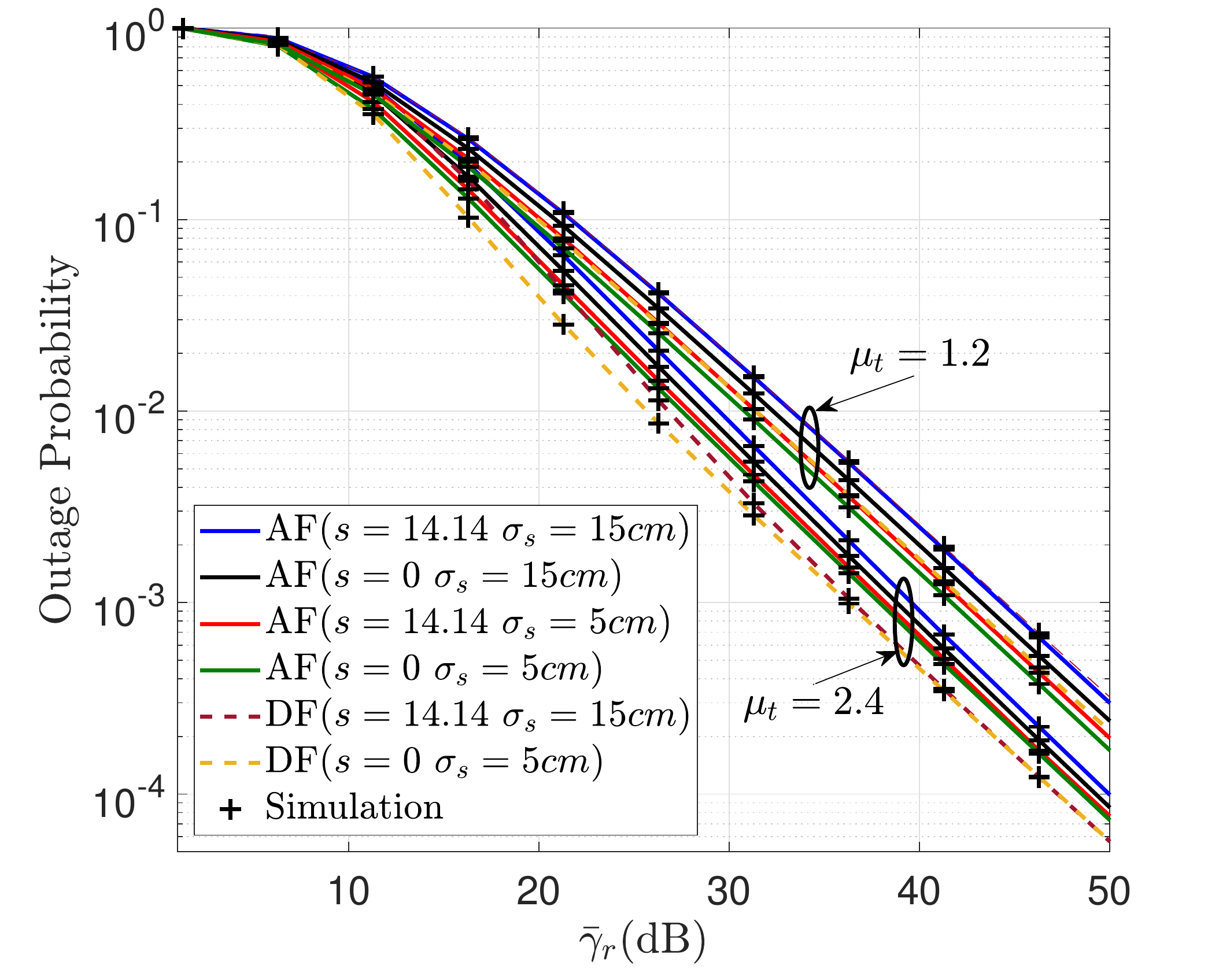}} 
	\subfigure[Average BER  at $\alpha_r=1.8$, $\mu_r=2$.]{\includegraphics[width=6cm, height=4.1cm]{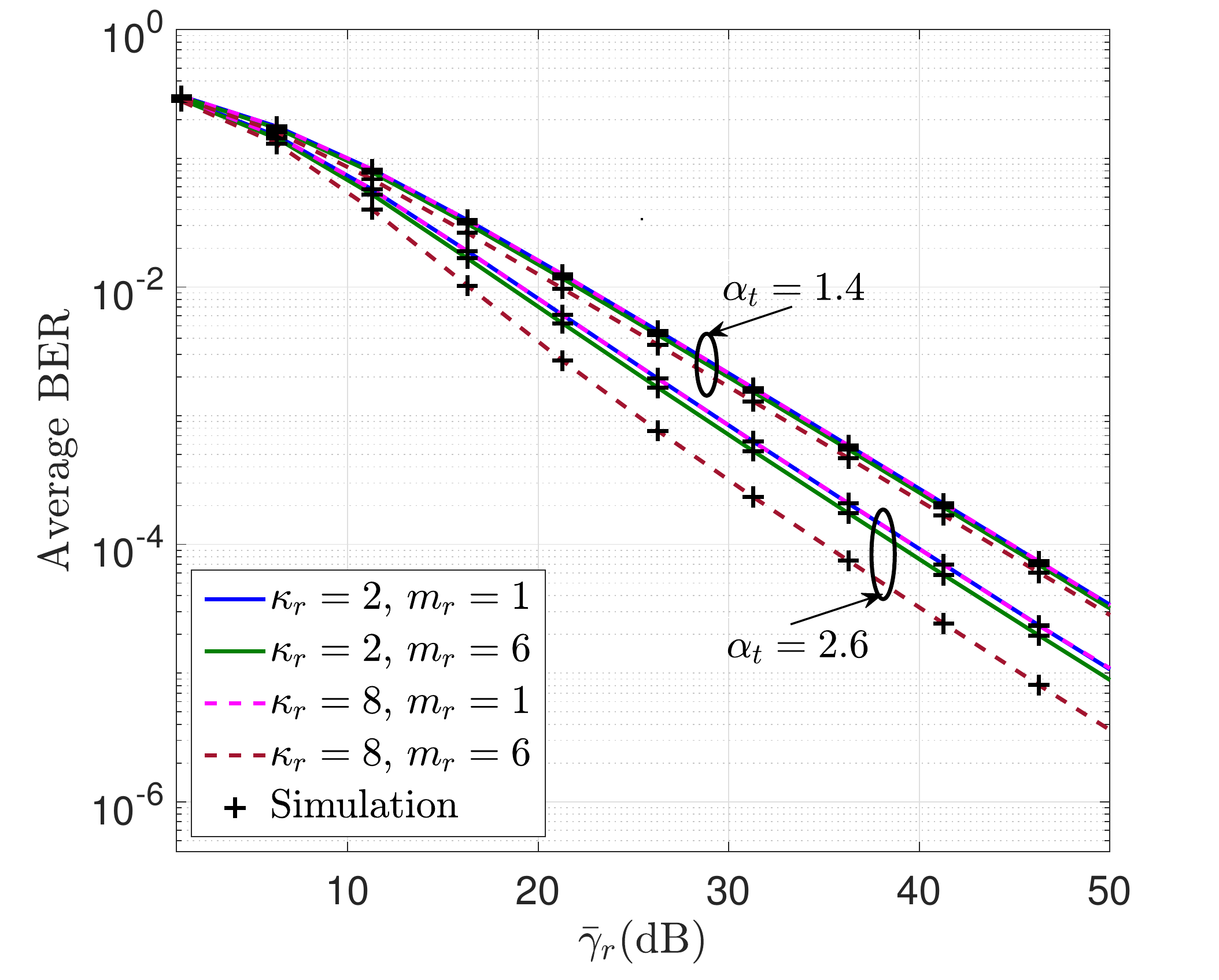}} 
		\subfigure[Ergodic capacity at $d_t=50$\mbox{m}. ]{\includegraphics[width=6cm, height=4.1cm]{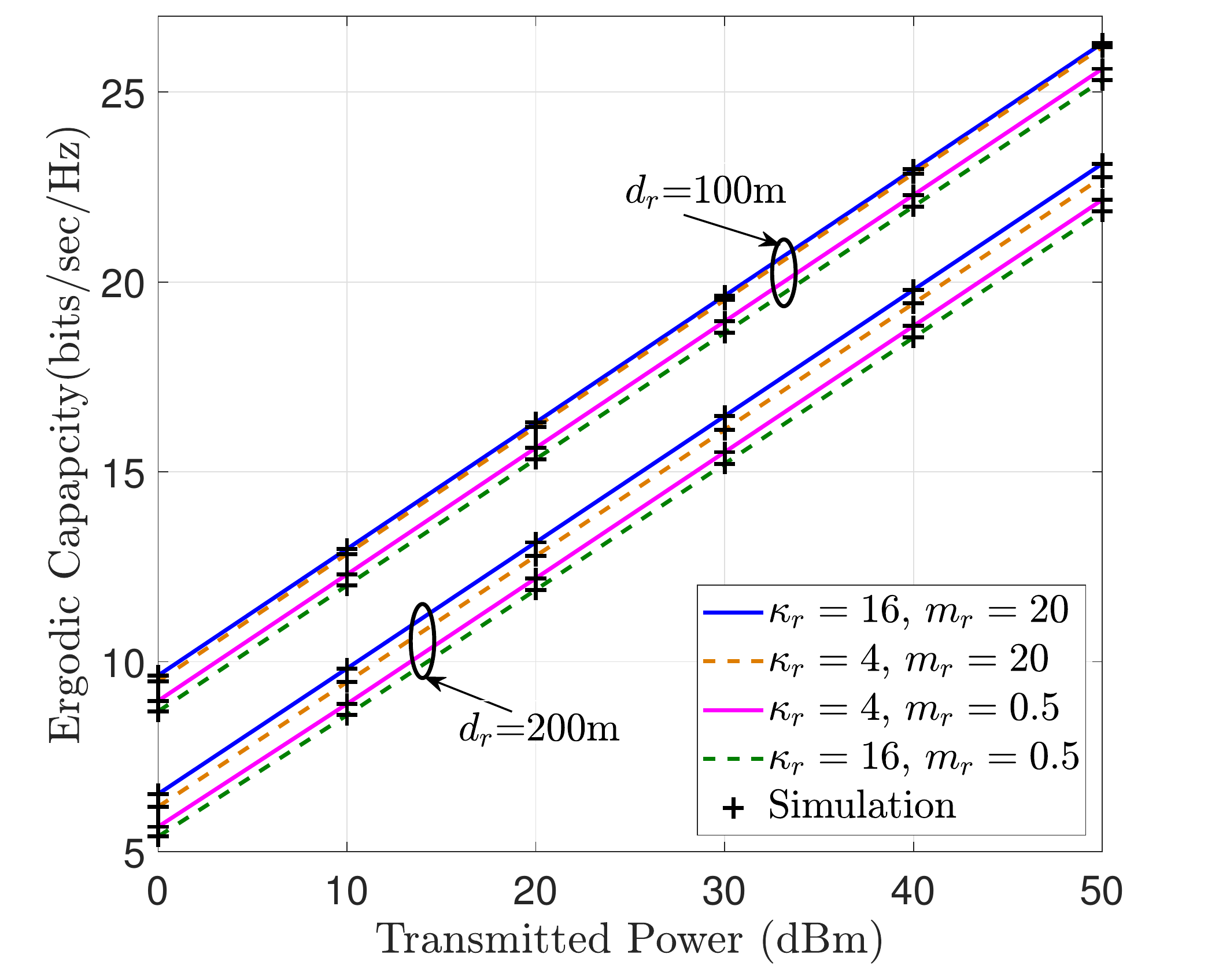}}
	\caption{Performance of fixed-gain relay-assisted RF-THz wireless link over mixed fading with non-zero pointing errors.}
	\label{fig:outage}
\end{figure*}

\section{Simulation Results and Discussions}\label{sec:sim_results}
We demonstrate  the performance of the considered RF-THz wireless system  and validate our derived analytical results using numerical analysis and Monte-Carlo simulations. To evaluate analytical expressions in \eqref{eq:cdf_fg}, \eqref{eq:ber_fg}, and \eqref{eq:capacity_fg},  we use MATLAB implementation of bivariate Fox's H function \cite{Illi_2017}, and take $10$ terms for the convergence of infinite series.  The  bivariate  Fox's H-function  requires the computation of  two contour integrals involving the ratio of the product of Gamma functions.  We also compare the proposed method with the existing DF relaying for the RF-THz system with zero-boresight pointing errors \cite{Boulogeorgos_Error,Pranay_2021_TVT}.	  To compute the path gain of the RF link with antenna gain $G_r=26$\mbox{dBi}, we use path loss  $L_r({\rm dB}) = 32.4+17.3\log_{10}(d_2)+20\log_{10} (10^{-9}f_r)$ \cite{Pranay_2021_TVT}, where $d_r$ is taken in  the range of $50$\mbox{m} to $200$\mbox{m}, and $f_r=6$\mbox{GHz} is the carrier frequency of the RF. We compute path gain of THz link as $H_{t} = \frac{cG_t}{4\pi f_t d_t} \exp(-\frac{1}{2}kd_t)$, where $G_t=55$\mbox{dBi}, $k=2.8\times10^{-4}$ is the absorption coefficient \cite{Boulogeorgos_Error}, $c$ is the speed of light, $f_t= 0.275$ \mbox{THz}, and $d_t=50$\mbox{m}. We use \cite{Yang2014} to compute the  parameters of pointing errors with $10$\mbox{cm}  antenna aperture  radius. A noise floor of $-170$\mbox{dBm/Hz} is taken for both THz and RF systems with $10$\mbox{GHz} and $20$\mbox{MHz} as the signal bandwidth for  THz and RF transmissions, respectively \cite{Sen_2020_Teranova}. 

First, we illustrate the impact of multipath clustering on the THz link (i.e., $\mu_t$) and the effect of non-zero boresight and jitter of pointing errors (i.e., $\sigma_s$ and $s$)  by plotting the outage probability versus  average SNR of the RF link $\bar{\gamma}_r$, as depicted in Fig.~\ref{fig:outage}(a). We consider $\alpha$-KMS parameters as $\{\alpha_r=1.8, \mu_r = 2, \kappa_r=4, m_r=2\}$. Fig.~\ref{fig:outage}(a) shows that the  outage probability improves with an increase in  $\mu_t$ since the multipath clustering enhances  channel conditions. Further, the figure shows that the effect of jitter is lesser at a higher $\mu_t=2.4$ and low RF average SNR but  incurs a penalty of almost $3$ \mbox{dB} if $\sigma_s$ is increased from $5$ \mbox{cm} to $15$\mbox{cm} at a lower $\mu_t=1.2$ and outage probability $10^{-3}$. It can also be seen from  Fig.~\ref{fig:outage}(a) that the non-zero values of  boresight incurs higher pointing errors degrading marginally the outage probability as compared to the zero-boresight.  The figure also shows that fixed-gain AF relaying performs close to the DF  without expensive decoding procedure and continuous monitoring of CSI in most of the scenarios.

Next, we depict the impact of non-linearity of the THz fading and shadowing effect of the RF link considering non-zero boresight parameter $s=14.14$ \mbox{m} and jitter $\sigma_s=5$ \mbox{cm} with $\mu_t=1.2$  on the average BER performance in Fig. \ref{fig:outage}(b). The figure shows a significant impact of the non-linearity factor $\alpha_t$ on the BER performance of the considered system. As such, the BER improves $10$ fold at a average SNR of  $40$ \mbox{dB} when the parameter $\alpha_t$ increases from $1.4$ to $2.6$ to get an average BER  of  $2 \times 10^{-5}$.  It can also be observed from the figure that the average BER performance degrades when the shadowing parameter increases from $m_r=6$ (less shadowing) to $m_r=1$ (severe shadowing). Further, with an increase in the parameter $\kappa_r=2$ to $\kappa_r=8$ at the given shadowing value, the average BER decreases marginally.

We demonstrate the impact of various parameters on the outage probability and average BER for a better insight into system performance. Note that  $\phi=37$ when $\sigma_s=5$ and \mbox{cm}) and $\phi=4.1184$ when $\sigma_s=15$ \mbox{cm} \cite{Yang2014}. When $\mu_t=1.2$, the outage diversity order is $0.9$  since $\alpha_t \mu_t=1.8$ is the minimum of $\alpha_r \mu_r=3.6$ and $\phi=4.1184$. Similarly, the diversity order becomes $1.8$ when $\mu_t=2.4$. Fig.~\ref{fig:outage}(a) shows that there is a change in the slope of plots corresponding to $\mu_t=1.2$ and $\mu_t=2.4$, but there is no change in the slope when pointing error parameters are changed. Thus, the diversity order analysis provides a design consideration to use high beam-width to compensate for the effect of pointing errors. Similar to the outage probability, the BER diversity order depends on the fading parameters of the THz link (i.e., $\alpha_t\mu_t= \{1.68, 3.12\})$ and becomes independent of pointing errors since $\phi(=37)> \mu_t\alpha_t$ and $\alpha_r\mu_r(=3.6)>\alpha_t\mu_t$. The plots confirm our analysis on the diversity order since there is a change in the slope of the plots in Fig \ref{fig:outage}(b) for different values of $\alpha_t$ and the slope does not change with the $\alpha$-KMS parameters $\kappa_r$ and $m_r$. Thus, the proposed analysis provides an insight into the deployment scenarios for the mixed RF-THz relaying considering various system and channel configurations.

Finally, Fig \ref{fig:outage}(c) presents the ergodic capacity performance for various RF link distance  $d_r=100$\mbox{m} and $d_r=200$\mbox{m} at a fixed THz link distance with $\alpha_t = 2, \mu_t = 2.2, \alpha_r=1.8, \mu_r = 2$, and $\sigma_s=10.6$ \mbox{cm}. This can be a typical situation for front-hauling with THz transmissions and broadband mobile access using the RF technology. Fig \ref{fig:outage}(c) shows that the ergodic capacity decreases with an increase in the RF link distance by $3$ \mbox{bits/sec/Hz} when $d_r$ is increased from $100$ \mbox{m} to $200$ \mbox{m}. Still, the ergodic capacity is significantly higher at $13$ \mbox{bits/sec/Hz} at a transmit power of $20$ \mbox{dBm} at $d_r= $200$ \mbox{m}$. The figure also shows the effect of shadowing $m_r$ and parameter $ \kappa_r $ (which is ratio of the total power of the dominant components and the total power of the scattered waves). It can be seen that the performance degrades with an increase in the shadowing effect (i.e., when $m_r$ decreases from $20$ to $0.5$) and a marginal change in the ergodic capacity with the parameter $\kappa_r$. 

We envision that the proposed RF-THz can be a viable alternative for the convergence of access networks and fronthaul/backhaul links over wireless technology. Consideration of co-channel interference and hardware impairment in the THz transmission link are a few possible directions for future works.

\bibliographystyle{IEEEtran}
\bibliography{bib_file_cl}
\end{document}